\documentclass[onecolumn]{aa}
\usepackage{graphicx}
\usepackage{txfonts}
\begin{document}

\title {
Production  of the Large Scale Superluminal
Ejections
of the Microquasar GRS 1915+105 by Violent Magnetic Reconnection
}
\author{E. M. de Gouveia Dal
Pino\inst{1}
\and
 A. Lazarian\inst{2}
}
\offprints{E.M. de Gouveia Dal Pino}
\institute{
Universidade de S\~ao Paulo, IAG, Rua do Mat\~ao 1226,
Cidade Universit\'aria, S\~ao Paulo 05508-900, Brazil\\
\email {dalpino@astro.iag.usp.br}
\and
Department of
Astronomy, University of Wisconsin, Madison, USA\\
\email {lazarian@astro.wisc.edu}
}

\date{Received ; accepted}

\abstract{

We here propose that  the large scale  superluminal ejections
observed in the galactic microquasar GRS 1915+105 during radio
flare events are produced by violent magnetic reconnection
episodes in the corona just above the inner edge of the magnetized
accretion disk that surrounds the central $\sim 10 \, M_{\odot}$
black hole. The process occurs when a large scale magnetic field
is established by turbulent dynamo in the inner disk region with a
ratio between the gas$+$radiation and the magnetic pressures
$\beta \simeq 1$, implying  a magnetic field  intensity  of $\sim
7 \times 10^8$ G. During this process, substantial angular
momentum is removed from the disk by the wind generated by the
vertical magnetic flux therefore increasing the disk mass
accretion to a value near (but below) the critical one ($\dot M \,
\sim  \, 10^{19} $ g s$^{-1}$). Part of the magnetic energy
released by reconnection heats the coronal gas ($T_c \lesssim 5
\times 10^8 $ K) that produces a steep, soft X-ray spectrum with
luminosity $L_X \simeq 10^{39} $ erg s$^{- 1}$, in consistency
with observations. The remaining magnetic energy released goes to
accelerate the particles to relativistic velocities ($v \,  \sim
\, v_A \, \sim \, c$, where $v_A$ is the Alfv\'en speed) in the
reconnection site through  first-order Fermi processes.  In this
context, two possible mechanisms have been examined which produce
 power-law electron distributions
 $N(E) \propto E^{-\alpha_E}$, with $\alpha_E = 5/2,\, 2$, and
 corresponding synchrotron radio power-law spectra with spectral
indices which are  compatible with that observed during the flares
($ S_{\nu} \propto \,\nu^{-0.75, - 0.5}$).

}

\titlerunning{Violent Magnetic Reconnection on Microquasar GRS 1915+105}
\authorrunning{de Gouveia Dal Pino and Lazarian}

\maketitle

\section{Introduction}
Galactic and extragalactic accretion-powered sources often exhibit
quasi-periodic
variability and ejection phenomena that may offer important clues to
the general physical processes in the inner regions of all classes of accreting
sources, from young stars to X-ray binaries and AGNs.

The X-ray source GRS 1915+105 was the first
galactic object to show evidence of
superluminal radio ejection (Mirabel \&
Rodriguez
1994, 1999) which is normally
interpreted as due to relativistic jets. At a
distance of $12.5 $ kpc and  probably with
a 10
solar mass black hole in the center of a binary system, this
microquasar offers an excellent
laboratory for the investigation of black
hole
accretion and associated jet
phenomena revealing, in time scales of
minutes to
months, a profusion of variable
emission structures in a large range of wavelengths (Dhawan, Mirabel, \&
Rodr\'iguez 2000).

From a detailed analysis of the observed RXTE X-ray variability of this source,
Belloni et al. (2000) have distinguished 12 different modes of variability that
they interpreted as transitions between three basic states: a
hard spectral-index state (denoted by C) with low X-ray luminosity, and two
states (denoted by A and B) with softer spectra that were modeled as thermalized
emission from an accretion disk, with state B being more luminous and hotter
than state A.
They also found that after a long steady hard class, the most frequent class of
variability was one characterized by a 30-minute cycle.

Variability also   has been observed in the radio and IR wavebands.
Mirabel et al. (1998) reported simultaneous X-ray, IR and
radio observations of these  30-minute cycles during which the
X-ray emission is
found to oscillate between states A/B and C. The IR/radio emission of the plasma
(which is consistent with synchrotron emission from an expanding nuclear jet;
see below) seems to be associated with the times of recovery from the X-ray dips
(see Figs. 2 and 3 from Mirabel et al. 1998), i.e., with the
times the source is changing from a period in state C to a  period in which it
rapidly oscillates between states A/B and C.

Large scale superluminal radio ejections have been also detected with MERLIN
emerging from the source when this was at the end of a 3-week extended phase of
low/hard X-ray state (also designated the
$plateau$ state; Fender et al. 1999; see also Fender et al. 2002).
Similarly, in a compilation of (VLBA) radio and
(RXTE
and BATSE) X-ray data, taken during several weeks, Dhawan,
Mirabel \& Rodriguez (2000; hereafter DMR)
have
distinguished two main states
of the system, a $plateau$ and a $flare$
state. The plateau state is
characterized by a flat radio spectrum
($S_{\nu}
\propto \nu^{0}$)  coming from
a compact region of  size  of a few AU, and
flux
density 1-100 mJy. During this
phase, the associated RXTE (2-12 keV) soft
X-ray
emission is weak, while the
BATSE (20-100 keV) hard X-rays are strong.
On the
other hand, during the flare
phase, optically thin ejecta are
superluminally
expelled up to thousands of AU with
fluxes up to 1 Jy at $\lambda 13$ cm
($S_{\nu} \propto \nu^{-0.6}$) with rise
time
$\lesssim  1$ day  and which  fade after
several days. The soft X-rays also flare
during
this phase and exhibit high
variability, while the hard X-rays fade for
few
days before recovering (see Figs. 1 and 4 of DMR).
The radio imaging over a range of
wavelengths (13,
3.6, 2.0, and 0.7 cm)
resolves the nucleus as a compact jet of
length
$\sim 10 \lambda_{cm}$ AU.
This nuclear jet varies in time scales of
$\sim
30$ minutes during minor X-ray/radio
bursts and reestablishes within
$\sim
18$ hours after a major
outburst. So far, GRS 1915 + 105 has been the only system among the galactic
accreting black holes of stellar mass where both AU-scale steady jets and large
scale superluminal ejections have been unambiguously observed.

It is generally believed that the X-ray
emission
of galactic black holes and
active galactic nuclei arises from hot gas
accretion flow or from an accretion-disk
corona. In the first class of models,
advection-dominated accretion flow (ADAF)
of disk material into the black hole (e.g.,
Narayan, Mahadevan, \& Quartaet 1997),
as well as, simultaneous
advection-dominated inflow-outflow (ADIOS) solutions
(Blandford
\& Begelman 1999) have been proposed. In the
second class, the presence of
magnetic fields adds
complexity
to the solutions but plays an essential role
in
the disk-coronal coupling, in the energy and angular momentum transport, and
in jet production and
acceleration (e.g., Liu,
Mineshige \& Shibata 2002).

The AU-scale nuclear jet of GRS 1915+105 developed in the plateau phase has
been
interpreted as synchrotron
emission from relativistic electrons
(e.g., Falke \& Biermann 1999), and
its origin has been attributed to disk
instabilities
in an ADAF-type flow by Belloni et al. (1997a,
b).
According to this model, the
30-minute soft X-ray variability observed
during
the plateau state, which involves a transition between states C and A/B, is
explained in
terms of periodic evacuation and refilling
of the
inner disk region on
time scales of seconds as a result of
thermal
viscosity. This process is believed
to occur in
the region where the radiation pressure
dominates
the gas pressure and the time
required to refill the disk with X-ray
emitting
hot gas is set by the thermal
viscous time scale. The observation of radio
oscillations with similar period
interval
supports the idea that the synchrotron
emission of
the nuclear jet is fed by
relativistic plasma injected from the inner
disk
during the dips in the soft X-
ray flux (DMR).

In contrast, the large superluminal radio
flares
observed in the $\sim $ 500 AU
scales cannot be explained by the same
viscous
disk instability model, because
they eject an order of magnitude more mass
than
the AU-scale $mini-jet$ and the
recovery time scale (of $\sim$ 18 hr) would
require according to that model a much
larger evacuation radius
of disk material well beyond the radiation
pressure-dominated region where the
instability occurs.

Recently, different groups have proposed alternative scenarios to explain the
variability of the
GRS 1915+105 microquasar. Livio, Pringle \& King (2003; see also King et al.
2004) have
suggested that the inner region of the accretion disk would switch between two
states.
In one of them (state A/B), the accretion energy would be dissipated locally to
produce
the observed disk luminosity, and in the other (state C) the accretion energy
would be
converted into magnetic energy and emitted in the form of a relativistic jet
(state C).
They have  attributed the  transition between the two states to dynamo
generation of a global
poloidal magnetic field. Similarly, Tagger et al. (2004),
 also  focusing mainly on the 30-minute cycles, have suggested that the
alternations
  between the high-soft and the low-hard states and the production of a nuclear
relativistic
  jet could be controlled by the successive accumulation and release of poloidal
magnetic flux
  (e.g., by magnetic reconnection) in the innermost parts of the accretion disk.
  They have  proposed a specific model to explain the quasi periodic variability
in
  the disk emission based on the development of the {\it accretion-ejection
instability}
  (Tagger \& Pellat 1999).  This  is a global mode formed by spiral waves that
travel back
  and forth in the inner regions of a disk threaded by a poloidal magnetic field
with a ratio
  $\beta$ between the gas pressure and the magnetic pressure close to unity. The
spiral waves
  become unstable by extracting energy and angular momentum from the inner
region
  (thus increasing the accretion) and transferring it outward.

Also, in a contemporaneous work, de Gouveia Dal Pino \& Lazarian (2004) have
suggested the possibility
that
the superluminal large scale radio flares of GRS 1915+105
could be produced by relativistic plasma
accelerated during violent magnetic reconnection
events occurring in the magnetized corona
just
above the inner regions of the accretion disk.
In the present work, we try to put together all the pieces of information above
in a consistent picture to show that violent magnetic reconnection could indeed
occur in the innermost parts of the disk and produce the large scale
superluminal ejections when $\beta$ approaches one and the mass accretion energy
becomes close to the critical value.

In
\P 2 we draw the potential scenario for
flare
production and the superluminal ejecta. Then
in \P
3, we try to quantify it by computing the
disk and
coronal parameters and evaluating the total
amount
of magnetic energy released during violent
reconnection in the corona; and in \P 4, we
evaluate both the electron and the
synchrotron
power-law spectra that is produced by the
accelerated particles through
first-order Fermi process in the
reconnection
site.
We
finally
conclude (in \P 5) with a brief discussion
of the
implications of our results.

\section {A Simple Possible Scenario}
\subsection{The BH magnetosphere}
Magnetized accretion disks around  rotating
(Kerr)
black holes (BHs) are
frequently invoked  to explain the high
energy
radiation  and jet
production and collimation in quasars and
microquasars. The magnetic field lines
originally frozen in the disk plasma will
deposit
along with accreting gas onto
the BH horizon, therefore developing a
magnetosphere around the later (see,
e.g., Wang,
Xiao, \& Lei 2002). Near the horizon,
no matter how chaotic may be the field threading the disk, the field
through the hole will become very $clean$ and ordered. If the disk tries to
deposit chaotic field on the hole, the field's closed loops will destroy
themselves on a time scale $ t \simeq R_H/c \sim 10^{-4}$ s $M_{10}$, leaving
the field ordered. If on the other hand, magnetic pressure from field lines
threading the disk temporarily push the hole's field into a clumped
configuration, it will spring back on this same time scale and make itself more
uniform (MacDonald et al. 1986).

The magnetic coupling (MC) of  a rotating black hole with its
surrounding disk has been recently investigated in some detail
(see e.g., Blandford 1999, Wang, Lei \&
Ma 2003, and references therein). The presence of some
closed field lines connecting the  BH with the inner disk edge (in a  variant of
the  Blandford-Znajek process, 1977), can lead to energy and
angular momentum transfer between the BH  and the disk if they are rotating at
different angular speeds. Though not a necessary condition, we will assume, for
simplicity,  in the
present study that the BH and the inner disk edge are nearly co-rotating so that
no significant angular
 momentum  and energy transfer is occurring between them.
The magnetic structure near the BH horizon will have a force-free
configuration determined by the MC process
and the strength of the field near the hole will be determined by the disk's
past history. The
net flux today must be equal to the total flux that the disk has deposited in
its entire past lifetime. These deposits will cause the net flux through the
hole to fluctuate stochastically. The rms field will grow with time until it
becomes so strong that it pushes its way back into the accretion disk and slows
down accretion. The resulting maximum strength is likely to be such as to
produce equilibrium between gas ram pressure and magnetic pressure in the
innermost parts of the disk (Wang, Lei and Ma 2003),
though Parker-Rayleigh-Taylor instabilities may
reduce this  strength (MacDonald
et al. 1986).

\subsection{The accretion disk and corona around the BH}
In the standard model for an accretion disk around a black hole, gravitational
energy is continually converted into kinetic, thermal, and magnetic energy by
the joint action of viscosity, and magnetic field line stretching and
reconnection. The viscosity can be mainly turbulent or magnetic, or both
(Shakura-Sunyaev
1973).

In the process of angular momentum transport in the disks, shear is
expected to convert  an existing poloidal magnetic field into a  toroidal
component. Then, probably
some of the toroidal flux will be used to generate new poloidal fields by
raising magnetic loops in a dynamo process.  Most of
the energy will be either dissipated within the
disk that will then emit thermalized radiation, or in the low density corona
above and below the disk that will radiate thermally or by Compton scattering,
or else the energy will be released in the form of a bulk flow, as a jet. In the
last case, it is possible that essentially all the accretion energy of the inner
disk is lost in the jet (see, Blandford \& Payne 1982,  Livio et al. 2003 and
references therein).

Let us assume as in previous works (e.g., Merloni 2003, Livio et al. 2003,
Tagger et al. 2004, de Gouveia Dal Pino \& Lazarian 2004) that the inner parts
of the accretion disk alternate between two states which are controlled by
changes in the global magnetic field.
Most of the magnetic field generated in a dynamo process in which the turbulence
is driven by, e.g., the magneto-rotational instability (Balbus \& Hawley 1991),
is a tangled small-scale magnetic field within the disk. As remarked by Livio et
al. 2003, as long as the magnetic field remains as such most of the dissipation
will occur locally within the disk that  will be radiative and very luminous.
However, if  a large scale poloidal magnetic field is generated then it is
possible that an energetic outflow can be driven.

Numerical simulations of the dynamo process in accretion disks (e.g., Kudoh et
al. 2002, Shibata 2003),
are still unable to
compute the generation of a global poloidal magnetic field,
{\it so that for the time being we must
rely on physical estimations rather than on detailed modeling} (Livio, Pringle
and
King 2003).
Tagger et al. (2004), for example, have suggested that
a gradual increase in the magnetic flux in the inner regions of the accretion
disk could be achieved by the combined action of the turbulent dynamo
associated,
e.g., with the magneto-rotational instability  (which would create a vertical
flux of opposite polarities in the inner and outer disk regions), followed by
advection of the magnetic flux with the gas to the inner disk regions. Though
magnetic diffusivity could interrupt this magnetic flux accumulation, there is
at least some observational evidence for it in the case of the magnetic field
structure of our own galactic disk that shows a strong  vertical field in the
inner tens of pcs of the galactic center which is almost three orders of
magnitude  more intense than the horizontal magnetic fields measured elsewhere
in the disk.
Livio (1997; see also Livio et al. 2003, King et al. 2004), in addition,
assuming a turbulent dynamo process with inverse cascade and reconnection of
magnetic loops above
and below the disk, have demonstrated that (provided that the distribution
$n(l)$
of the lengths of the  magnetic loops satisfies a relation $n(l) \propto l^{-
3/2}$) large scale poloidal fields can be generated such
that $L_{j} \sim L_{acc}$, where $L_{j}$ is the jet power and $L_{acc}$ is the
disk accretion
luminosity.

Considering the aspects above,
we assume here that during the $plateau$ state that precedes a radio flare, a
large scale poloidal field is progressively built in the disk by a turbulent
dynamo process. The action of buoyancy forces will also make the disk unstable
against the Parker-Rayleigh-Taylor instability and horizontal magnetic field
lines will raise from the disk forming large scale loops in the rarified hot
corona.
We further assume that once the dynamo process establishes a global poloidal
field over a substantial region of the disk, it will be able to maintain that
field for a period of time. The length of time the source spends in this state
is hard to estimate without a complete physical model, however observations
indicate that the system remains in the plateau state for prolonged periods of
time.

The vertical field flux will give rise to a wind that will remove  angular
momentum from the disk, therefore significantly increasing the accretion rate
(possibly at a rate greater than the rate due to the disk viscosity). Also, with
the  accumulation (trap) of vertical flux in the inner regions the ratio between
the gas+radiation pressure to the magnetic field pressure ($\beta$) will soon
decrease to one.
Under these conditions, events of reconnection of magnetic field lines with
opposite polarization will be inevitable, and in the innermost regions this
process may become eventually very violent when enough magnetic energy is
stored
in the corona. We show below that this occurs when  $\beta \sim 1$ and the
accretion rate approaches the critical value.  We then argue that this could
explain the large scale radio flares in GRS 1915+105 with the ejection of the
observed relativistic blobs from the innermost regions.


\subsection{Magnetic reconnection site in the inner disk-corona region}

Considering the assumptions above, the resulting qualitative
structure of the magnetosphere of the hole and the accretion disk
must be like that shown in Figure 1 (which was adapted from
Macdonald \& Thorne 1982, and Macdonald et al. 1986). In the inner
disk region (of radius $R_X$), there is a site which may be
appropriate for violent magnetic reconnection. Surfaces of null
poloidal field lines mediate the geometry of the open field lines
anchored into  the BH horizon with the opened lines of the disk
wind  and those connecting the disk with the BH horizon.
Labeled as "$Y$ $netral$ $zone$" in Figure 1, these magnetic null
surfaces begin or end on $Y$ points. Across each  null surface,
the poloidal field suffers a sharp reversal of direction.
According to the Amp\`ere's law, large electric currents must flow
out of the plane shown in Figure 1, along the null surfaces, and
in the presence of finite electric resistivity, dissipation of
these currents will lead to reconnection of the oppositely
directed field lines (e.g., Biskamp 1997, Lazarian \& Vishniac
1999).  We here  investigate the possibility that the magnetic
energy released by reconnection near the Y point region (see also
Figure 2) is able to accelerate  the plasma to relativistic
velocities and produce the synchrotron radio flares (see below).

\begin{figure}
\centering
\includegraphics[width=9cm]{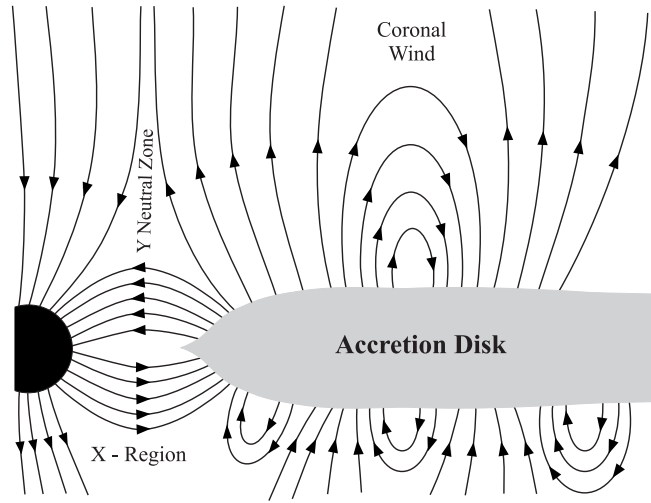}
\caption{Schematic drawing of the magnetic field geometry in the
inner disk region at $R_X$. The acceleration occurs in the
magnetic reconnection site at the   Y type neutral zone (see the
text). }
\end{figure}

The release of magnetic energy by reconnection is  believed to be
the most promising mechanism for production of solar flares.  {\bf
Both direct and indirect astrophysical evidence show that the
nature seems to know the way of producing fast reconnection in
spite of the theoretical difficulties involved (see, e.g.,
Lazarian, Vishniac \& Cho 2004).} Y-type neutral zones are also
present, for example, in eruptive magnetic arcades of the solar
corona, like those associated to two-ribbon solar flares (see,
e.g., Takasaki et al. 2004, Sakajiri et al. 2004). In these cases,
a possible interpretation of the observations is that a current
sheet forms between an emerging magnetic flux and the pre-existing
magnetic flux of opposite polarization and magnetic reconnection
in this zone will cause the formation of a filament/plasmoid. This
will temporally inhibit further reconnection and establish a
temporally stable magnetic configuration. During this stage, there
will be considerable magnetic energy storage due to the continuing
converging motion of the two fluxes. Eventually an eruption
develops (for example, from sheared or twisted arcades lying
underneath the magnetic null point), the filament/plasmoid is then
ejected and a large quantity of magnetic energy is released (see,
e.g., Takasaki et al. 2004).

{\bf Lazarian \& Vishniac (1999) proposed a mechanism at which
flares can naturally emerge  as a result of an instability
mediated by turbulence. In their model the reconnection rate
increases due to magnetic field wandering caused by turbulence.
Fluxes of non-parallel magnetic field lines reconnect very slowly
if the turbulent velocities are much smaller than the Alfven
velocity. However, outflows caused by reconnection increase the
level of turbulence and therefore the reconnection rate. As a
result of such a positive feedback a flare develops.}

Whatever is the particular mechanism, the energy released by
magnetic reconnection
 is  expected to ultimately drive violent
outward motions in the surrounding plasma that can produce copious
amounts of solar cosmic rays via direct acceleration along the
open magnetic field lines or via a first order Fermi process
(e.g., Reames 1995). Similar mechanisms have been recently
proposed to accelerate the plasma in magnetic reconnection zones
of young born accretion-induced collapse pulsars to produce
ultra-high energy cosmic rays (de Gouveia Dal Pino \& Lazarian
2000, 2001).

The several stages described above for magnetic reconnection and
solar flare production may  be  also operating in our accretion
disk-corona-BH system, but since in the present study it is not
crucial to follow the detailed evolution of the magnetic
configuration and of the features that develop during the
reconnection process, we will consider a  simpler scenario.
According to the discussion in the previous sections, we argue
that whenever $\beta$ becomes small enough and the accretion rate
attains values near the critical one in the inner disk region,
then the lines of opposite polarization near the Y-zone
 will be pressed together   by the accreted material sufficiently
rapidly as to undergo violent reconnection events. Energy stored
in the magnetic field will be released suddenly to heat the plasma
and accelerate charged particles. Some of the high speed particles
will travel downward to the BH or to the disk to the foot points
of the magnetic-arches, yielding bright X-ray emission, others
will spew outward giving rise to the relativistic radio ejecta
(see below). \footnote {In the scenario just described, the high
accretion rate may also cause the compression and shear of
magnetic loops that may exist in the lower parts of the inner disk
corona,  under the Y-zone dome. These could also undergo violent
reconnection and  even produce plasmoids that could eventually
erupt and provoke particle acceleration as well, like in the Sun
(see discussion in section 4). }

\begin{figure}
\centering
\includegraphics[width=8cm]{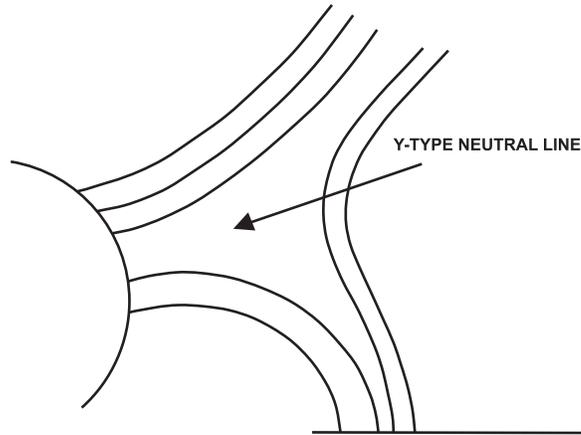}
\caption{ A closer view of the Y-type neutral zone where the field
lines coming from the central source encounter those opened by the
disk wind and those trapped by the funnel inflow emanating from
the $R_X$ region to form an Y point of null magnetic field.}
\end{figure}

We notice that the merging of the closed field lines of the loops and the open
field
lines in the disk further out from the inner radius can also lead to
reconnection, however, we find that the
amount of magnetic energy released in these regions is not sufficient to
accelerate the particles
to produce the observed radio luminosities in the large flares (see below).

For reconnection to occur in the inner disk region $R_X$, near the
Y point, the poloidal flux from  the disk and the flux anchored in
the central magnetic structure surrounding the BH must have
opposite signs, like  in Figure 1. As pointed out  by Tagger et
al. (2004), this suggests that large ejections with violent flares
will occur only when the system is in that configuration. Violent
reconnection will not take place in the periods when both fluxes
have the same sign. This dichotomy is observed for example in the
interaction between the terrestrial and the solar wind magnetic
field. Now, the magnetic flux trapped in the BH horizon is a sum
of all magnetic flux accreted during the whole history of the
source. If the magnetic flux in the disk is either produced by a
turbulent dynamo in the disk itself or advected from a companion
star, its sign is expected to change from time to time, although
the characteristic time scale  of this variation is difficult to
compute in the absence of a detailed modelling of the dynamo in
the disk or in the companion star. On the long term, a sufficient
number of reconnection events might in turn cancel the flux
trapped in the BH horizon and change its sign as well. In the case
of the solar dynamo, it is not clear yet why the field reversals
take place every 22 year. In the case of GRO 1915$+$105, Tagger et
al. (2004) have suggested that this long term evolution of the
field configuration could be of the order of one to few  years so
that part of the time the field fluxes in the inner disk edge and
in the hole would be antiparallel and the rest  of the time
parallel.

As remarked above, observations of the Sun indicate that the
amplification, dissipation and reconnection of the magnetic field
lines is likely to occur not steadily, but sporadically, involving
violent flares followed by more gradual recoveries to the
pre-flare magnetic field configuration with no obvious
periodicity. Similar behavior, i.e., violent flares followed by
more slowly recovering of the pre-flare state  is also detected in
the observed superluminal radio ejecta of GRS 1915+105, thus
suggesting that magnetic reconnection could be also associated to
the production of relativistic blob ejections that rapidly move to
the $\sim 100$ AU scales. In the next section, we try to quantify
the ideas just described.

We notice that after reconnection, the partial destruction of the magnetic flux
in the inner disk will make $\beta$ to increase and make the disk resumes the
less magnetized condition of the plateau state with most of the energy being
dissipated locally within the disk instead of in the outflow.

\section{Building the Model for the Superluminal Ejections During the Flares}

We
adopt the field geometry
of the magnetized accretion
disk/corona as described in the previous section
(Fig. 1).

We could first try to make an order-of-magnitude estimate of the amount of
magnetic
energy that could be extracted from the inner disk region by
simply assuming that it is approximately equal
to the disk accretion
energy rate $G M_{\star}\dot M/R_X$.
For a BH with mass
$M_{\star} = 10 M_{\odot } = M_{10}$ and
Schwartzschild
radius
$R_{\star}  = 2 G M_{\star}/c^2 = 2.96 \times 10^6 \, M_{10}$
cm, this results
\begin{equation}
\dot W_B  \, \simeq  \, 10^{39} \,   {\rm
erg \,  s}^{-
1} \,
M_{10} \, \dot M_{19} \, R_{X,7}^{-1}
\end{equation}
\noindent
where we have assumed that the  inner radius
of the
accretion disk ($R_X$) corresponds
approximately
to the last stable orbit of the
BH, $R_X \simeq \, 3 R_{\star} \, \simeq \,
10^7 $
cm, and that the accretion rate
$\dot M \lesssim \dot M_{Edd}$, where
$\dot M_{Edd} = 1.9 \times 10^{19} \,
M_{10}$ g s$^{-1}$ is the Eddington
critical accretion rate (Shakura \& Sunyaev 1973),
$\dot M_{19}$ is $\dot M$ in units of
$10^{19}$ g s$^{-1}$, and
$R_{X,7}$ is $R_X$ in units of $10^7$ cm.
This
estimate
of $\dot W_B$ can be
compared to   the
observed X-ray luminosity of GRS 1915+105
($\sim
10^{39}$ erg  s$^{-1}$).

Now, in order to evaluate more precisely the amount of magnetic
energy that can be extracted through violent reconnection from the
Y neutral zone to produce superluminal ejecta, we need first to
evaluate the physical conditions both in the magnetized disk and
corona.

\subsection{Disk and Coronal Parameters}

To evaluate the disk quantities at $R_X$, we
may
employ the standard disk model
(Shakura \& Sunyaev 1973; Frank et al.
1992).
In the inner disk region (named the
X-region), the radiation pressure is larger
than
the thermal pressure and
the disk will be $radiation$
pressure-dominated.
In this case,

\begin{equation}
T_d   \, \simeq  \, 1.2 \times 10^{8} \,   {\rm
K} \,
 \alpha_{0.1}^{-1/4} \,
M_{10}^{1/4} \,
R_{X,7}^{-3/4}
\end{equation}
\noindent
and
\begin{equation}
n_d   \, \simeq  \, 2.8 \times 10^{17} \,   {\rm
cm}^{-3}
\, \alpha_{0.1}^{-1} \,
M_{10}^{-1/2} \, \dot M_{19}^{-2} \,
R_{X,7}^{3/2} \, q_{0.82}^{-8}
\end{equation}
\noindent
where,
$q \, = \,[1 \, - \, (R_{\star}/R_X)^{1/2}
]^{1/4} \, = \,
0.82$,
$T_d$ and $n_d$  are the disk gas number
density
and temperature, respectively,
$\alpha_{0.1}$ is the Shakura-Sunyaev
viscous
coefficient in units of 0.1 (e.g., Frank et al. 1992), and
$\dot M_{19}$ is the disk accretion rate in units of $10^{19}$
g s$^{-1}$.

The width $H$ of the disk
in the radiation pressure-dominated
region
is given by
\begin{equation}
\frac{H} {R_X} \, \simeq \, 0.7 \, \dot
M_{19} \,
R_{X,7}^{-1} \, q_{0.82}^4
\label{H}
\end{equation}

As remarked in the previous section, the violent magnetic reconnection events
that may originate large scale relativistic ejections are expected to occur when
the large scale magnetic field in the inner disk region, which is generated by
dynamo process, attains an intensity as large as
$B_d^2/8\pi \simeq P_d$, where $P_d$ is the disk pressure which is dominated by
the radiation pressure, i.e.,
when $\beta \simeq 1$, giving

\begin{equation}
B_d \, \simeq \, 6.9 \times 10^8 \, {\rm G}
\,
\alpha_{0.1}^{-1/2} \,
M_{10}^{1/4} \, R_{X,7}^{-3/4}
\label{BX}
\end{equation}

Recently, Liu, Mineshige, \& Shibata (2002),
have
proposed a simple model to
quantify the parameters of  a magnetized
disk-corona system. Assuming, like in
the solar corona, that gas evaporation at
the foot
point of a magnetic flux tube
quickly builds up the density of the corona
to a
certain value, and the tube  radiates away
the
heating due to magnetic reconnection through
Compton
scattering,
they derived the following coronal
quantities (as
a function of disk quantities) for a
radiation
pressure-dominated disk

\begin{equation}
n_c   \, \simeq  \, 7.6 \times 10^{14}  \, {\rm
cm}^{-3}
\, \alpha_{0.1}^{-15/16} \,
\beta_{1}^{-3/4} \,
M_{10}^{-25/32} \, \dot M_{19}^{-2} \,
R_{X,7}^{75/32} \, l_8^{-3/4}
\end{equation}
\noindent
and

\begin{equation}
T_c   \, \simeq  \, 4.9 \times 10^{8} \,  {\rm K}
\,
\alpha_{0.1}^{-15/32} \,
\beta_{1}^{-3/8} \,
M_{10}^{-25/64} \, \dot M_{19}^{-1} \,
R_{X,7}^{75/64} \, l_8^{1/8}
\end{equation} \noindent
where  $n_c$  and $T_c$ are the coronal gas number density and
temperature, respectively, $\beta_{1}$ is the value of $\beta$ in
the disk in units of 1, and $l_{8}$ is the scale height of the  Y
neutral zone in the corona in units of $10^8$ cm. Although the
quantities above have been computed for the purposes of the
present model only in  the inner disk radius $R_X$, they are
actually valid for any radius of the disk-corona within the
radiation pressure-dominated region. We note  that, according to
the Liu et al.'s model, smaller values of $\beta$ would result
more energy dissipation through the corona and larger coronal
temperature and density, as indicated by Eqs. (6) and (7).

\subsection{Rate of Magnetic Energy Release
in the Y-type neutral zone}

Let us assume that the poloidal magnetic field that raises in the
corona just above the inner disk region ($B_X$) is of the order of
the local disk magnetic field,
 $B_d$ (eq. 5), that has been produced and pushed to the inner disk region as
described
 in the previous section (see also, e.g., Arons 1993, de Gouveia Dal Pino \&
Lazarian 2001).
\footnote{
 We notice that this value is a little larger than the one estimated
 by assuming, e.g.,  that the ejected gas takes the bulk of
the accretion power, i.e., $L_j \sim L_{acc}$  (where
$L_j \simeq B_c^2 R_X^2 v_{\phi}/2$,
 $L_{acc} \simeq \dot M v_{\phi}^2$, and
$v_{\phi}$ is the Keplerian toroidal velocity in the disk; see also Livio et al.
2003).}

The rate of magnetic energy that can be extracted from the Y-zone
in the corona (above and below the disk) through reconnection is
$\dot W_B \, \simeq \, (B_X^2/ 8\pi) \, \xi v_{A} \, (4\pi R_X
)\,\L_X$, where $\xi = v_{rec}/v_A$ is the reconnection efficiency
factor, $v_{rec}$ is the reconnection velocity, $v_A = B_X/ (4 \pi
n_c  m_p)^{1/2}$ is the coronal Alfv\'en speed, $m_p$ is the
hydrogen mass,
and $L_X$ is the length of the reconnection
region which according to Figure 3
is
$L_X/\Delta R_X\approx v_A/\xi v_A \approx
\xi^{-
1}$.
 For the conditions here investigated we
find that
$v_A \, \simeq \, c$. Substitution of these
relations into the equation above
gives
\begin{equation}
\dot W_B  \, \simeq  \,  \frac{ B_X^2 } {8
\pi}  c
\, (4\pi R_X )\, \Delta R_X
\end{equation}

In order to estimate $\Delta R_X$ (or $L_X$), we need to model the
reconnection process itself. Classical reconnection schemes use to
provide too slow reconnection rates. However, there is ample
observational evidence that reconnection in Astrophysics may take
place at very large rates which are comparable with the Alfv\'en
speed (see e.g., Dere 1996). In the framework of solar flares,
considerable progress has been made in recent years towards the
determination of more realistic reconnection rates during flare
events based mostly on the separation speed of $H\alpha$ flare
ribbons (e.g., Qiu et al. 2004; Jing et al. 2005; Sakajiri et al.
2004, Takasaki et al. 2004). These observations indicate that the
reconnection speed can be as high as few tenths of the Alfv\'en
velocity (Takasaki et al. 2004). Since in the conditions we deal
with in the present analysis this speed approaches $c$, the
corresponding reconnection rate can be rather large.

As mentioned earlier, in the turbulent reconnection model proposed
by Lazarian \& Vishniac (1999, henceforth LV99; 2000, see also
Lazarian 2005 for an extensive review on this subject) wandering
of magnetic field lines was suggested as the ultimate cause of
fast reconnection. {\bf An alternative model (Bisckamp, Schwarz \&
Drake, 1997, Shay et al. 2003) attempts to explain fast
reconnection appealing to plasma properties, such as anomalous
resistivity (Parker 1979). These models do not exclude each other.
For instance, while dealing with anomalous resistivity in most
astrophysical situations one has also to deal with the effects of
turbulence. At the same time the reconnection rates in the LV99
model may get an initial boost from the anomalous resistive
effects. This boost may be important when the initial turbulent
velocities are much smaller than the Alfven speed. In our present
analysis, we focus our attention on regions of strong magnetic
fields, like those around the BH in the X-region during nearly
critical accretion rate. For those regions the resistivity should
be anomalous and the fast reconnection should be guaranteed (with
$v_{rec} \sim v_A$).}

Following LV99, we can estimate the width of the current sheet for
which the resistivity should be anomalous:
\begin{equation}
\Delta R_X =\frac{c\Delta B}{4\pi n_c Z e
v_{th,c}}
\end{equation}
\noindent
where $\Delta B \simeq 2B_X$ denotes the
change of
the magnetic field across
the reconnection region, and $v_{th,c} =
(\frac {5/3
\, k T_c} {m_p})^{1/2}$ the
thermal velocity of the ions of charge $Ze$
in the
corona
\begin{equation}
v_{th,c}   \, =  \, 2.65 \times 10^{8}   {\rm
cm \,
s}^{-1} \, \alpha_{0.1}^{-15/64}
\, \beta_{1}^{-3/16} \,
M_{10}^{-25/128} \, \dot M_{19}^{-1/2} \,
 R_{X,7}^{75/128} \, l_8^{1/16}.
\end{equation}

Substitution of Eqs. (5), (6), and (10) into
Eq. (9) results
\begin{equation}
\left(\frac {\Delta
R_X } {R_X} \right)
\, \simeq \,   \, 0.0034 \, Z^{-1}\,
\, \alpha_{0.1}^{43/64}
\, \beta_{1}^{15/16} \,
M_{10}^{157/128} \, \dot M_{19}^{5/2} \,
R_{X,7}^{-599/128} \, l_8^{11/16}
\end{equation}

\begin{figure}
\centering
\includegraphics[width=11cm]{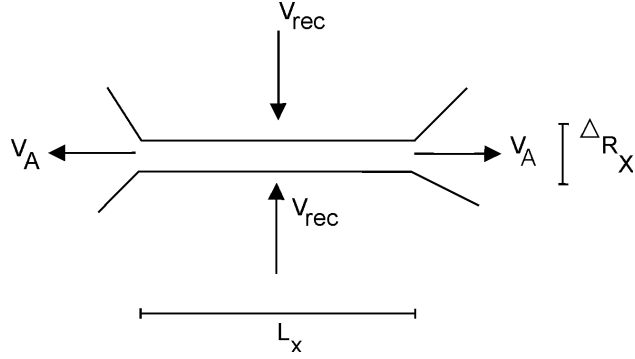}
\caption{ Schematic representation
of a reconnection region.}
\end{figure}

\noindent

Now, if we plug  the result above and
Eq. (5) into
Eq. (8), we get
\begin{equation}
\dot W_B \, \simeq  \, 2.4 \times 10^{39}
{\rm
erg \, s}^{-1} \,
\alpha_{0.1}^{-1} \,
M_{10}^{1/2} \, R_{X,7}^{1/2} \, \left(
\frac {\Delta
R_X/R_X} {0.0034} \right)
\end{equation}

\noindent
and the corresponding reconnection time is

\begin{equation}
t_{rec}  \, \simeq  \, \frac {R_X} {\xi v_A} \, \simeq  \,
3.3 \times 10^{-4}   \,
{\rm s}^{-1} \,  \xi^{-1} \, R_{X,7}
\end{equation}
\noindent
which indicates that the release of magnetic energy is very fast.

The analysis
of Eqs. (11) and (12) shows that the
adoption of a somewhat smaller viscosity parameter
$\alpha$ (which would imply a more quiescent disk),
by an order of magnitude, for instance,  would require a disk accretion rate
only few times larger in order to keep the values of
$\Delta R_X/R_X$ and  $\dot W_B$ unaltered.

Eq. (12) gives  the total expected amount of magnetic energy rate
released by fast reconnection in the Y-zone during the flare state
of GRS 1915+105. Part of this energy will heat the gas that will
produce X-ray luminosity, and the remaining will accelerate the
charged particles to relativistic velocities producing a violent
radio ejecta. But what is the fraction that goes to each part, and
what is the potential acceleration mechanism in the reconnection
site that can produce the superluminal radio ejections?

The  evaluation of  the exact fraction of the magnetic energy that
goes to each of the processes above is a task out of the scope of
the present work, as it depends on the detailed modelling of the
conditions of the disk-corona and of the efficiency of each
process. However, assuming that most of the magnetic energy
released through reconnection in the corona goes to heat the
plasma, Liu et al. (2002) have computed the average value of the
Compton parameter ($y$) for an optically thin corona with an
original  distribution of thermal electrons, $y= \frac {4 k T_c}
{m_e c^2} n_c \sigma_T l$, which was weighted over all the disk
radii. According to their results, for a disk near the critical
accretion ($ \dot M \sim \dot M_{Edd}$), $y \, < \,  0.8$, which
implies a soft (and steep)  X-ray spectrum, just as required by
GRS 1915+105 observations  during the  flare state, with a
spectral index $\alpha_{r-x} \, = \left({ \frac{9} {4} + \frac {4}
{y} }\right)^{1/2} - \frac {3} {2} >  \, 1.2$. This result is
qualitatively consistent with the  present model, as it predicts,
for a magnetized disk-corona a predominantly soft X-ray spectrum
during nearly critical accretion.

Since observations do indicate that both,
heating
and acceleration of the plasma to
relativistic
velocities are occurring, as suggested by the observed disk X-ray
luminosity of the order of $\sim 10^{39}$ erg s$^{-1}$,
and by the radio flares which in turn
suggest a jet mechanical power also of the order
of
$\sim 10^{39}$ erg s$^{-1}$  at least in the nuclear region (see, e.g, Falcke \& Biermann
1999),
we will here
assume that
a substantial fraction  of
$\dot W_B$ (Eq. 12) goes to
accelerate
the particles.

Let us now try to answer the second part of
the
question above, i.e., what could be the
acceleration mechanism in the reconnection
region able  to produce the relativistic
ejecta
and the radio flare?

\section{ The Acceleration Mechanism during
the
Flare State}

As stressed by de Gouveia Dal Pino \&
Lazarian
(2001), the particular mechanism of particle
acceleration
during reconnection events is still unclear
in
spite of
numerous attempts
to solve the problem
(see
LaRosa et al 1996,  Litvinenko 1996).
Cosmic rays from the Sun confirm that the
process
is sufficiently
efficient in spite of the apparent
theoretical
difficulties for its
explanation.

Particles could, in principle,  be accelerated at once to the
relativistic  energies by the large induced electric field within
the reconnection region (e.g., Bruhwiller \& Zweibel 1992,
Haswell, Tajima, \& Sakai 1992, Litvinenko 1996; de Gouveia Dal
Pino \& Lazarian 2000, 2001), but one-shot acceleration mechanism
can make it difficulty to explain  how particles will afterwards
produce a power-law particle and synchrotron spectrum. Another
possibility is that particles are accelerated by a Fermi process
in the reconnection site. In this regard, we discuss below two
viable sorts of this process.

A schematic representation of a reconnection region is shown in
Figure ~3. The upper and lower parts of the magnetic flux move
towards each other with the velocity $v_{rec}$. As a result,
charged particles rays in the upper part of the reconnection zone
"see" the lower part of the magnetic flux to approach them with a
velocity $2v_{rec}$. In this case an acceleration process
analogous to the first-order Fermi acceleration of cosmic rays in
magnetized shocks may take place (e.g., Longair 1992) and we can
write
\begin{equation}
\frac{\Delta E}{E}=\frac{2
v_{rec}}{c}\cos\theta
\end{equation}
For a particle probability distribution
$p(\theta)=2\sin\theta \cos\theta
d\theta$,
the average energy gain per crossing of the
reconnection region is
\footnote{ We note that since the particle
Larmour
radius around the magnetic field in the
reconnection zone,
$r_B = \gamma m_e c^2/ (Ze B_X) \, \simeq \,
9 \times 10^{-
6}$ cm $ \gamma_{3.5} \, B_{X,7} \,
< < \,  \Delta R_X$
(where $\gamma_{3.5} \,  = \, \gamma/3.5$ is the Lorentz
factor
for electrons  with velocities $\sim 0.92 \,
c$),
the particles will remain confined time
enough in
the reconnection zone to be efficiently
accelerated through the Fermi mechanism
before
escaping from the system.}

\begin{equation}
\langle\frac{\Delta
E}{E}\rangle=\frac{v_{rec}}{c}\int^{\pi/2}_{
0}
2\cos^2\theta \sin\theta
d\theta=\frac{4}{3}\frac{v_{rec}}{c}
\end{equation}
where $\theta$ is the pitch angle between
the
particle velocity and the magnetic
field.
The round trip of the particles  between the
upper
and lower magnetic
fluxes will produce an increment of energy
twice
as large. Thus the
ratio of the particle energy, $E$, after
such a
round trip to its
original energy, $E_0$, will be
\begin{equation}
\delta=\frac{E}{E_0}=1+\frac{8}{3}\frac{v_{re
c}}{c}
\end{equation}
Now,  assuming that the escape probability of a particle from the
acceleration zone is similar to the one computed  for a shock
front, $4v_{rec}/c$,
  then probability
to stay within the reconnection region is simply
\begin{equation}
P=1-\frac{4v_{rec}}{c},
\end{equation}
and the electron particle spectrum is
obtained
from (see e.g., Longair
1992)
\begin{equation}
N(E)dE\sim E^{-1+\frac{lnP}{ln\delta}}dE \propto E^{-5/2}dE
\label{spect}
\end{equation}

\noindent This equation indicates that the particle spectrum
produced by first-order Fermi acceleration within  the
reconnection site is steeper than that normally produced in shocks
(for which a similar simplified treatment that ignores losses and
non-linearities provides $N(E)\propto E^{- 2}$), so that the two
processes may be, in principle, distinguished. Now, the
Synchrotron radiation from electrons with a power-law distribution
$N(E) \, \propto \, E^{-p}$, has a power-law frequency radio
spectrum $S_{\nu} \propto \, \nu^{-\alpha_r}$, where the spectral
index is given by $\alpha_{r} \, = \, (p \, - \, 1)/2 \, \simeq \,
0.75$. Though crude, this  estimated value is consistent with the
observed radio spectrum during the flares ($\alpha_{r} \simeq
0.6$).

Finally, the particle acceleration mechanism discussed above does
not disregard the possibility that the relativistic fluid may be
also produced behind shocks. As in the Sun, plasmoids formed by
reconnection of the  field lines lines near the Y-zone
   may violently erupt and cause
the formation of a shock front. This raises the alternative
possibility of a first-order Fermi acceleration process of the
ionized gas behind the shock front and the development of a
power-law spectrum with particle spectral index of the order of
two ($N(E)\propto E^{- 2}$) as remarked above and, as such, a
corresponding synchrotron spectrum with a spectral index also of
the order of the observed one, $S_{\nu} \propto \, \nu^{-0.5}$.

After a radio flare, while the system is
slowly
recovering its original
magnetic field configuration, the plateau
phase must
resume. During this recovering
phase, the magnetic field intensity in the
X-region will be smaller and also the
mass accretion rate.
In this
state, the X-ray spectrum is
observed to be hard
and  flat.  As we have seen in \P 3.2, the
model of Lin et al. (2002) predicts
a harder X-ray spectrum when the accretion
rate in the magnetized  disk is very sub-Eddington for which a spectral index
$\alpha_{r-x} \, <  \,
1.2$ should be expected.

\section{Conclusions and Discussion}

We have here investigated the origin of  the large scale  superluminal ejections
observed in
the galactic microquasar GRS 1915+105 during radio flare events
and proposed a very simple model in which
they could
be
due to violent magnetic reconnection episodes in
the
corona just above the inner edge of the
magnetized accretion disk
that surrounds the central $\sim 10 \, M_{\odot}$ black hole.  In this region,
when a large scale magnetic field is built by a turbulent dynamo process, and
there is approximate  balance between the disk magnetic and gas$+$radiation
pressures ($\beta \simeq 1$), which implies  a strong magnetic
field  intensity   $B_X \simeq
7 \times 10^8$ G, reconnection between the magnetic lines that arise from the
disk with those
of the magnetosphere of the BH may become violent. During this regime,
the vertical magnetic flux generates a wind that will substantially remove
angular momentum from the disk therefore increasing
the disk mass
accretion to a value near the critical one
(with $\dot M \, \sim  \, 10^{19} $ g s$^{-1}$).

Whenever the system reaches the condition above and the magnetic
field lines of the BH magnetosphere and those of the disk have
opposite polarization, it may undergo an episode of violent
reconnection that lasts for very short time $\sim 3 \times
10^{-4}$ s. A substantial part of the total magnetic energy
released by reconnection ($\sim 2 \times 10^{39}$ erg s$^{-1}$)
heats the coronal gas to $T_c \simeq 5 \times 10^8 $ K that
produces a steep, soft X-ray spectrum with the expected luminosity
from observations. Assuming  that the remaining released magnetic
energy goes to accelerate the particles to relativistic velocities
($v \,  \sim v_A \, \sim \, c$, where $v_A$ is the Alfv\'en speed)
in the reconnection site through  first-order Fermi processes, we
have found that a power-law electron distribution  $N(E) \propto
 E^{-5/2, \, -2}$, and a synchrotron radio power-law spectrum
($ S_{\nu} \propto \, \nu^{-0.75, -0.5}$) are produced with radio
spectral indices which are comparable to the one observed during
the flares.

After each  violent reconnection  event, the system must resume a sub-critical
disk mass accretion with smaller magnetic field intensity, and a slower
reconnection rate,
thus recovering the more quiescent plateau state, as described in \P 2 to 4.

Although until the present,  GRS 1915 + 105 has been the only system among the
galactic accreting black holes where large scale ($\gtrsim $ 100 AU)
superluminal
ejections have been observed (maybe because this system is the only one
that meets the
necessary conditions as investigated here),
the magnetized-disk-corona model here discussed
could be applicable, in principle, also to other X-ray
binary sources, or even to
the microquasar extragalactic counterparts, i.e., quasars with evidence of
superluminal ejection.

\acknowledgements This paper has benefited from many valuable
comments from an anonymous referee. E.M.G.D.P. has been partially
supported by grants of the Brazilian Agencies FAPESP (grant
1997/13084-3) and CNPq.


\begin{thebibliography}{}
\bibitem[]{ }
Arons, J. 1986, in Plasma Penetration into
Magnetospheres, eds. N. Kylafis, J.
Papamastorakis, and J. Ventura (Iraklion:
Crete
Univ. Press), 115

\bibitem[] { }
Arons, J.  1993, \apj, 408, 160

\bibitem[] { }
Belloni,  T., M\'endez, M., King, A. R., van den Klis, M. , \& van Paradus, J.
1997a, \apj, 479, L145

\bibitem[] { }
Belloni,  T., M\'endez, M., King, A. R., van den Klis, M. , \& van Paradus, J.
1997b, \apj, 488, L109

\bibitem[] { }
Biskamp, D., Schwarz, E., \& Drake, J.F.
1997,
Phys. Plasmas, 4, 1002

\bibitem[] { }
Biskamp, D. 1997, in Advanced Topics in
Astrophysical and Space Plasmas, eds.
E.M. de Gouveia Dal Pino, A. Perat, G.A.
Medina
Tanco, and A.C.L. Chian (The
Netherlands: Kluwer), 165

\bibitem[]{}
Blandford, R.D. 1999, in ASP Conf. Sers., vol. 160, Astrophysical Discs, eds.
J.A. Selwood \& J. Goodman (San Francisco: ASP), p. 265

\bibitem[] { }
Blandford, R.D., \& Begelman, M. C. 1999, MNRAS, 303, L1

\bibitem[] { }
Blandford, R.D., \& Znajek, R.L. 1977, MNRAS, 179, 433

\bibitem[] { }
Bruhwiller, \& Zweibel 1992

\bibitem[] { }
de Gouveia Dal Pino, E.M. \& Lazarian, A.
2000,
\apj, 536, L31

\bibitem[] { }
de Gouveia Dal Pino, E.M. \& Lazarian, A.
2001,
\apj, 560, 358

\bibitem[] { }
de Gouveia Dal Pino, E.M., \& Lazarian, A. 2004, in proc. X Marcel Grossman
Conference (in press)

\bibitem[] { }
Dere, K.P. 1996, ApJ, 472, 864

\bibitem[] { }
Dhawan, V., Mirabel, I. F., \& Rodrig  ez, L. F.
2000,
\apj, 543, 373

\bibitem[]{}
Falcke, H., \& Biermann, P. L. 1999, A\&A, 342, 49

\bibitem[] { }
Fender, R.P., Garrington, S.T., McKay, D.J., et al. 1999, MNRAS, 304, 865

\bibitem[] { }
Fender, R.P., Rayner, B., Trushkin, S.A., et al. 2002, MNRAS, 330, 212

\bibitem[] { }
Frank, J. King, A., \& Derek, R. 1992, in Accretion Power in Astrophysics,
(Cambridge: Cambridge Univ. Press)

\bibitem[] { }
Gosh, P., \& Lamb, F.K. 1978, \apj, 223, L83

\bibitem[] { }
Kudoh, T., Matsumoto, R., \& Shibata, K. 2002, PASJ, 54, 267

\bibitem[] { }
Haswell, C. A., Tajima, T., \& Sakai, J.-L. 1992, \apj, 401, 495

\bibitem[] { }
Jing, J., Qiu, J., Lin, J, Qu, M., Xu, Y., \& Wang, H. 2005, \apj,
620, 1085

\bibitem[] { }
King, A.R., Pringle, J.E., West, R.G., and Livio, M. 2004, MNRAS (in press)
\bibitem[]{}

LaRosa, T.N., Moore, R.L., Miller, J.A., \&
Shore, S.N. 1996, ApJ, 467, 454

\bibitem[] { }
Lazarian, A. 2005, in  Magnetic Fields in the Universe: from
Laboratory and Stars to Primordial Structures, American Institute
of Physics Procs. (NY), eds. E.M. de Gouveia Dal Pino, G. Lugones
and A. Lazarian), in press

\bibitem[] { }
Lazarian, A. \& Vishniac, E. 1999, \apj, , 517, 700

\bibitem[] { }
Lazarian, A. \& Vishniac, E. 2000, Rev. Mex.
Astron. Astrof., 9, 55

\bibitem[] { }
Lazarian, A.  Vishniac, E., \& Cho, J, \apj, 603,  180

\bibitem[]{}
Litvinenko, Y.E. 1996, ApJ, 462, 997

\bibitem[]{}
Liu, B. F., Mineshige, S., \& Shibata, K. 2002, astro-ph/0205257

\bibitem[] { }
Livio, M., Pringle, J.E., \& King, A.R., 2003, \apj, 593, 184

\bibitem[] { }
Livio, M. 1997, in IAU Colloq. 163, Accretion Phenomena and Related Outflows,
ed. D.T. Wickramasinghe, G.V. Bicknell, \& L. Ferrario (ASP Conf. Ser. 121; San
Francisco: ASP), 845

\bibitem[]{}
MacDonald, D.A., \& Thorne, K.S. 1982, MNRAS, 198, 345

\bibitem[]{}
MacDonald, D.A.,  Thorne, K.S., Price, R. H., \& Zhang, X.-H.1986, in Black
Holes: The Membrane Paradigm, p. 120

\bibitem[]{}
Mirabel, I.F., \& Rodriguez, L. F. 1994, Nature, 371, 46

\bibitem[] { }
Merloni, A. 2003, MNRAS, 341, 1051

\bibitem[]{}
Mirabel, I.F., \& Rodriguez, L. F. 1999, ARA\&A, 37, 409

\bibitem[]{}
Narayan, R., Mahadevan, R., \& Quataert, E. 1998, in The Theory of Black Hole
Accretion Disks, ed. M. A. Abramowicks, G. Bjornsson, \& Pringle, J. E.
(Cambridge: Cambridge Univ. Press), 148

\bibitem[]{}
Qiu, J., Wang, H., Cheng, C. Z., \& Gary, D. E. 2004,  \apj, 604,
900

\bibitem[] { }
Reames, D.V. 1995, Rev. Geophys., 33
(suppl.), 585

\bibitem[] { }
Sakajiri, T., Brooks, D. H., Yamamoto, T., Shiota, D., Isobe, H.,
Akiyama, S., Ueno, S., Kitai, R., \& Shibata, K. 2004, \apj, 616,
578

\bibitem[] { }
Shakura, N. I., \& Sunyaev, R. A. 1973, A\&A, 24, 337

\bibitem[] { }
Shibata, K., and Aoki, S. 2003, astro-ph/0303253




\bibitem[] { }
Tagger, M., Varni\'ere, P., Rodriguez, J., et al. 2004, \apj,607, 410

\bibitem[] { }
Tagger, M., \& Pellat, R. 1999, A\&A, 349, 1003

\bibitem[] { }
Takasaki, H., Asai A., Kiyohara, J., Shimojo, M., Terasawa, T.,
Takei, Y., \& Shibata, K. 2004, \apj, 613, 592

\bibitem[] { }
Vishniac, E. , \& Lazarian, A. 1998, \apj,
511, 193

\bibitem[]{}
Wang, D.X., \& Lei, W. H., \& Ma R.-Y. 2003, MNRAS, 342, 851

\bibitem[]{}
Wang, D.X., Xiao, K., \& Lei, W. H. 2002, MNRAS, 335, 655
\end{thebibliography}
\end{document}